\documentstyle[osa,aps,prb,epsfig]{revtex}
\begin{document}
\twocolumn[\hsize\textwidth\columnwidth\hsize\csname
@twocolumnfalse\endcsname
\title{$^{\ast\ast\ast}$
Temperature Dependence of Optical Spectra for the CDW Gap and
Optical Phonons in BaBiO$_3$}
\author{$^{\ast}$Hyun-Tak Kim, Hidetoshi Minami and Hiromoto Uwe}
\address{Institute of Applied Physics, University of Tsukuba, Ibaraki 305, Japan}
\maketitle{}
%\newpage
\begin{abstract}
We have studied the temperature dependence of the optical spectra
for the CDW gap and optical phonons in a single crystal of
BaBiO$_3$ and a film deposited by laser ablation. The integrated
optical conductivity for optical phonons decreases rapidly with
increasing temperature above 150 K, while the single-particle
excitation across the CDW gap remains constant.\\ \\
\end{abstract}
]
%\newpage
%\narrowtext
%\section{INTRODUCTION}
BaBiO$_3$ with a perovskite structure has been known as a mother
material for the superconducting Ba$_{1-x}$K$_x$BiO$_3$ (BKBO),
Ba$_{1-x}$Rb$_x$BiO$_3$ (BRBO) of $T_c$=30K at x=0.4 and
BaPb$_{1-x}$Bi$_x$O$_3$ (BPBO) of $T_c$=13K at x=0.25. The
transition from semiconductor to superconductor occurs with
doping. The semiconductivity of BaBiO$_3$ is worth studying in
order to elucidate the mechanism of their superconductivity.

BaBiO$_3$ is a semiconductor despite the metallic prediction of
the band-structure calculations in the cubic phase.$^1$ In order
to explain this energy gap, Charge-Density-Wave (CDW) theory has
been introduced. The CDW instability originates from the aversion
of Bi to the 4$^+$ valence, which leads to a disproportiontion of
Bi$^{3+}$ (6s$^2$) and Bi$^{5+}$(6s$^0$) valences on alternate
sites. A commensurate CDW with alternately expanded and contracted
Bi-O bond lengths on adjacent sites has been found. In the
measurement of neutron and X-ray diffraction, Cox and Sleight
found that the breathing mode distortion occurs in the real
material.$^2$

Experiments to probe the electronic structure of BaBiO$_3$ have
been performed by several groups, principally using optical
spectroscopy techniques. The energy gap of CDW of 2 eV and
four-optical phonons in a far-infrared region were
revealed.$^{3,4}$ Also, an indirect gap corresponding to the
activation energy; the 0.24 eV was found in mid-infrared
absorption spectra.$^5$

In this paper, we investigate temperature dependences of optical
phonons in a far-infrared region and the CDW gap of 2 eV for a
single crystals of BaBiO$_3$ prepared by a flux method and for
thin films deposited by laser ablation, respectively.

A single crystal of BaBiO$_3$ was prepared by a flux method from a
mixture of BaO and Bi$_2$O$_3$ powders. In a furnace, the
temperature was increased to 1015$^{\circ}$C over 8 hours. The
temperature was kept at 1015$^{\circ}$C for 30 minutes, decreased
for 15 hours at a rate of 1$^{\circ}$C/h, and then allowed to
decrease to room temperature naturally. Then, single crystals with
a sufficiently large area appeared at the surface of the bowl with
a golden color. Their crystal structures were found to be
monoclinic by means of X-ray diffraction (XRD).

The film were synthesized by laser ablation at the substrate
temperature of 600$^{\circ}$C at the laser-energy density of 2
J/cm$^2$ under an oxygen atmosphere at the pressure of 200 mTorr
on the transparent (100)MgO substrate. The thickness of the film
was about 280 nm. The laser-ablation system and deposition
procedures of the film were explained in a previous paper$^6$.

The reflectances and transmittances of BaBiO$_3$ were observed by
using spectrometers with a visible near-infrared and of the
Fourier Transformer of Infrared (FTIR). Reflectances were observed
in the infrared range of 600 cm$^{-1}$ (80 meV) from 30 K to 300
K. Transmittances in a visible near-infrared region were observed
from 20$^{\circ}$C to 300$^{\circ}$C and examined no heat
detriment at 50$^{\circ}$C after heating.

Figure 1 (a)-(g) shows optical conductivities, ${\sigma_{op}}$, of
optical phonons observed in the range of 50-600 cm$^{-1}$ (50-80
meV) at various temperatures. These conductivities were calculated
from reflectances by the Kramers-Kronig relation. The
conductivities decrease with increasing and decreasing temperature
from 150 K. Each peak position of phonons does not significantly
vary with temperature. By the relation of the sum rule
(${\int}{\sigma_{op}}dE$=$he^2N_{op}/2m$), we can define the
electronic density, $N_{op}$; here, the rest mass of electron is
used for a mass, $m$. The $N_{op}$ is shown in Fig. 2 as a
function of temperature, which shows a maximum at 150 K. Since the
conventional optical phonons should not exhibit temperature
dependence of optical conductivity; this anomalous temperature
dependence is attributed to the optical phason proposed by M. J.
Rice$^7$. The each phonon decreases $\sigma_{op}$ rapidly with
temperature above 150 K, as shown in Fig. 3 (a)-(d), here, the
phonons are called the phonon 1, 2, 3 and 4 with decreasing order
of the wave number. The decrease of the optical conductivity
indicates that the electronic density of CDW decreases. The
decreased electron density to the electron density of 150 K nearly
agrees with the carrier density deduced from the hall effect$^8$.
This exhibits that the electronic condensate in the CDW state
loose electrons excited to the conduction band, in which the
optical spectrum would be broad unobserable. Probably, the
decrease of the optical conductivity below 150 K may be due to a
structural-phase transition. In the case of 1D-metal such as KCP
and K$_{0.3}$MoO$_3$, peaks in a far-infrared region have been
known as the pinned acoustic phason should be searched below 50
cm$^{-1}$.

Figure 4 (a)-(e) show the transmittances observed near 16000
cm$^{-1}$ (CDW gap of 2 eV) at various temperatures.  The
BaBiO$_3$ film deposited by laser ablation was used for the
transmittance observation. A thickness of the film was about 280
nm. The transmittances at 17000 cm$^{-1}$ increase with increasing
temperature. The minimum positions (17000 cm$^{-1}$) of each
transmittance do not vary with temperature. A damage at the film
occurs above 350$^{\circ}$C. The transmittance observed at
50$^{\circ}$C after the heating is nearly the same as that
observed at room temperature, as shown in Fig. 4 (e), which
indicates that the film is cooled without a damage. The
transmittance shows a large conductivities at 20$^{\circ}$C and
300$^{\circ}$C were calculated from absorption coefficients
calculated from Gauss-Newton numerical analysis, as shown in Fig.
5. The conductivities decrease with increasing temperature at the
maximum position. However, the total integrated $\sigma_{op}$
remains rather constant all through the temperatures.

In conclusion, we have found the temperature-dependent optical
conductivity for phason. This is characteristic of the optical
phason in the CDW state of BaBiO$_3$.

%\section*{ACKNOWLEDGEMENTS}

\newpage

\begin{figure}
\vspace{-1.0cm}
\centerline{\epsfysize=18.0cm\epsfxsize=8.0cm\epsfbox{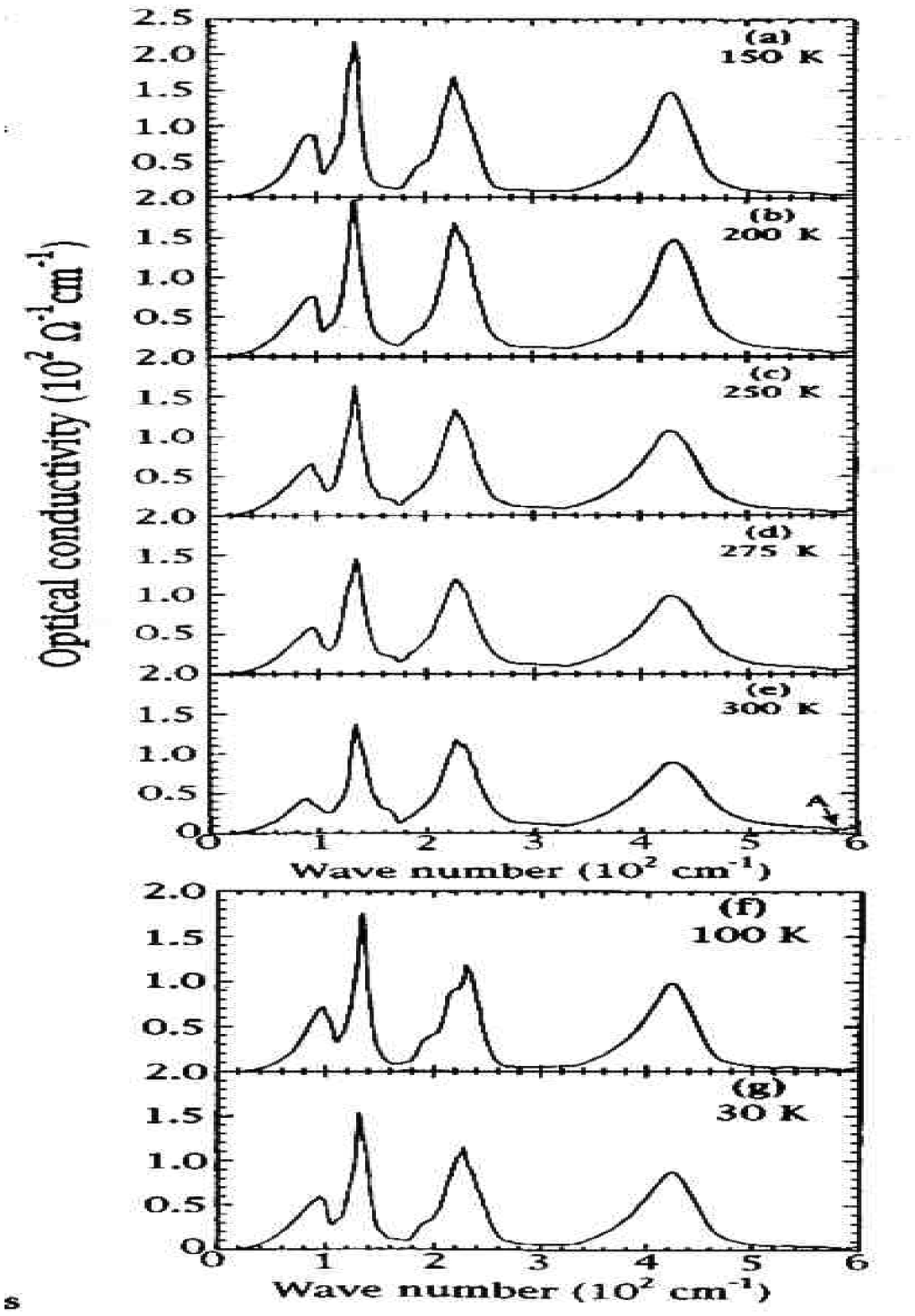}}
\vspace{-0.5cm} \caption{Temperature dependence of optical phonons
for BaBiO$_3$}
\end{figure}

\begin{figure}
\vspace{-0.5cm}
\centerline{\epsfysize=10.0cm\epsfxsize=8.0cm\epsfbox{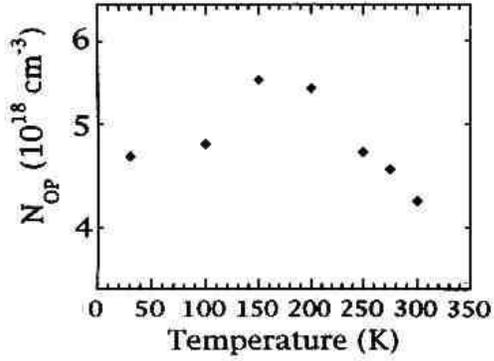}}
\vspace{-2.0cm} \caption{Temperature dependence of electronic
density, $N_{op}$.}
\end{figure}

\begin{figure}
\vspace{-1.0cm}
\centerline{\epsfysize=13.0cm\epsfxsize=8.0cm\epsfbox{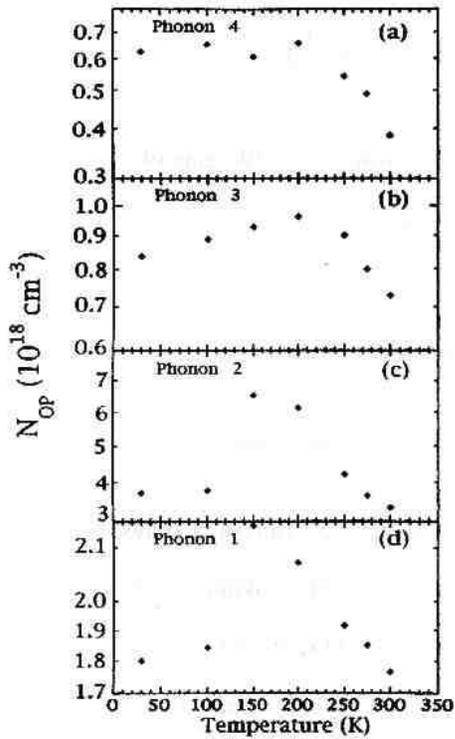}}
\vspace{-1.0cm} \caption{Temperature dependence of each one of
phonons.}
\end{figure}

\begin{figure}
\vspace{-0.5cm}
\centerline{\epsfysize=13.0cm\epsfxsize=8.0cm\epsfbox{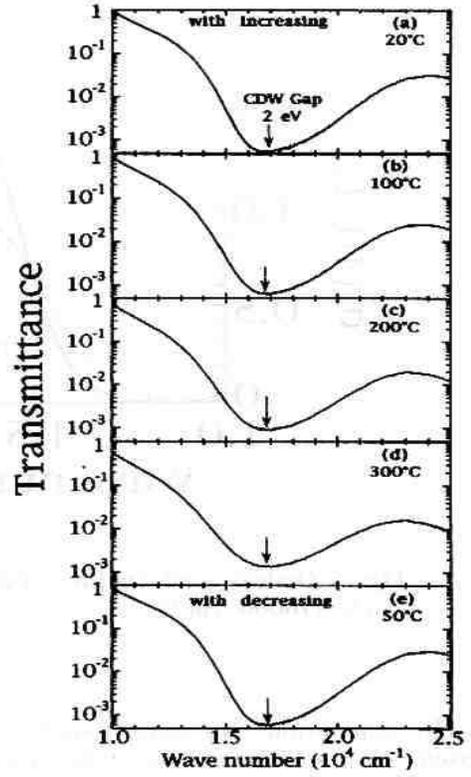}}
\vspace{-0.0cm} \caption{Temperature dependence of transmittance
near the CDW gap of 2 eV (16000 cm$^{-1}$).}
\end{figure}

\begin{figure}
\vspace{-1.0cm}
\centerline{\epsfysize=9.0cm\epsfxsize=8.0cm\epsfbox{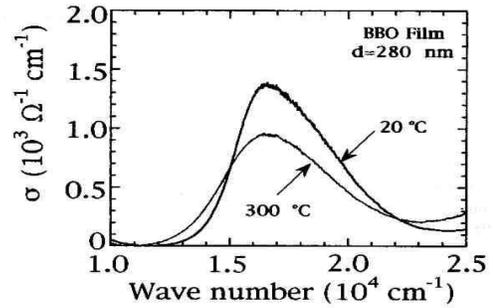}}
\vspace{-1.5cm} \caption{Optical conductivities at 20$^{\circ}$
and 300$^{\circ}$C near the CDW gap of 2 eV (16000 cm$^{-1}$).}
\end{figure}

\end{document}